\documentclass[%
 aip,
cp,  
 amsmath,amssymb,
 reprint,%
]{revtex4-2}

\usepackage{graphicx}
\usepackage{dcolumn}
\usepackage{bm}

\usepackage[utf8]{inputenc}
\usepackage[T1]{fontenc}
\usepackage{mathptmx} 
\def\p{\bm{p}}
\def\q{\bm{q}}
\def\u{\bm{u}}
\def\v{\bm{v}}
\def\r{\bm{r}}
\def\w{\bm{w}}

\begin{document}

\title{Integrable systems on symmetric spaces from a quadratic pencil of Lax operators }

\author{Rossen Ivanov} 
 \email[Corresponding author: ]{rossen.ivanov@tudublin.ie}
\affiliation{School of Mathematics and Statistics,
  Technological University Dublin,
  Grangegorman Lower,
  Dublin D07 ADY7, Ireland
}


\date{\today} 

\begin{abstract}
The article surveys the recent results on integrable systems arising from quadratic pencil of Lax operator $L$, with values in a Hermitian symmetric space. The counterpart operator $M$ in the Lax pair defines positive, negative and rational flows. The results are illustrated with examples from the A.III symmetric space. The modeling aspect of the arising higher order nonlinear {S}chr{\"{o}}dinger equations is briefly discussed.
\end{abstract}

\maketitle

  This work is dedicated to Professor Vladimir Gerdjikov on the occasion of his 75th birthday.  

\section{\label{sec:level1}Introduction}

\subsection{NLS models on symmetric spaces}

The multi-component generalisations of the nonlinear {S}chr{\"{o}}dinger equation (NLS) are heavily studied in soliton theory. A large class of matrix generalizations involve Lax pairs taking values in some simple Lie algebra. The simple Lie algebras admit splitting which is associated to the structure of a Hermitian symmetric space.  Details about the classification of the Hermitian symmetric spaces can be found in the classical book of Helgason \cite{Helgason}. The seminal works of Fordy, Kulish and Athorne \cite{KuFo,AtFo} marked the beginning of the practical use of symmetric spaces in the soliton theory. Since then integrable systems on finite dimensional Lie algebras and their symmetric spaces have been studied considerably in the literature \cite{GG1, GGI, GGK,VSG-basic, NK-VG, VSG1, GGIK, I04,V13, GI21, GMV, GMV2, GGMV,GGMV11}.

 We begin with a brief description of the necessary algebraic concepts. 
 
 A simple Lie algebra $\frak{g}$ over the complex numbers admits the splitting \begin{align}\label{spl}
  \mathfrak {g}= \mathfrak{k}\oplus \mathfrak{m}\, ,
\end{align}
  where $\mathfrak{k}$ is a subalgebra of $\mathfrak{g},$ and $\mathfrak{m}$ is the complementary subspace of $\mathfrak{k}$ in $\mathfrak{g}.$
  In addition, \begin{align}
        [\mathfrak{k}, \mathfrak{k}] \subset \mathfrak{k}\, , \qquad [\mathfrak{k},\mathfrak{m} ] \subset  \mathfrak{m}\, , \qquad [\mathfrak{m},\mathfrak{m}] \subset \mathfrak{k}\, .
    \label{sym-relation}
\end{align}   Denoting by $K$ and $G $ the Lie groups, associated to $\mathfrak{k} $ and $\mathfrak{g}$ correspondingly, the linear subspace $\mathfrak{m} $ is  identified with the tangent space of $G/K,$ which is used as a notation for the corresponding symmetric space.

\subsection{Hermitian symmetric spaces}

The Hermitian symmetric spaces are a special class of symmetric spaces for which there is a special element $J\in \mathfrak{k} $ such that
\begin{equation}\label{Jel}
  \mathfrak{k}=\{ X \in \mathfrak{g } : \, \, [J,X]=0    \}, \quad [J, \mathfrak{k}]=0.
\end{equation}
   It is clear then that the Cartan subalgebra $\mathfrak{h}\subset \mathfrak{k}\subset \mathfrak{g},$ and the element $J$ can be chosen from the Cartan subalgebra, $J \in \mathfrak{h}.$ In other words, $J$ will be chosen to be diagonal. 
  Furthermore, this element is highly degenerate, in a sense that  $\mathrm{ad}_J,$ which is an $n \times n$ matrix ($n=\mathrm{dim}(\mathfrak{g})$) has only three eigenvalues; $0$ and $\pm a,$ and the subspace $\mathfrak{m}$ can be split
further as
\begin{align*}
  \mathfrak{ m}= \mathfrak{m^+}\oplus \mathfrak{m^-}, \quad  \mathfrak{m^{\pm}}=\{X^{\pm}: \,\, [J, X^{\pm}]=\pm a X^{\pm} \}.
\end{align*}

    \subsection{Outline of the NLS-type models}

The NLS-type equations on symmetric spaces introduced by Fordy and Kulish originate from a spectral problem
    \begin{equation}\label{Lax1}
    \begin{split}
   &  i \psi_x=(\lambda J +Q(x,t))\psi \equiv L(\lambda) \psi,\\
  &  i \psi_t=P(x,t,\lambda )\psi \equiv M(\lambda) \psi, \qquad Q(x,t) \in \mathfrak{ m}, \quad P(x,t,\lambda)\in \mathfrak{ g}.
  \end{split}
\end{equation}
Clearly, we have the natural decomposition    $P= P_{\mathfrak{ k}}(x,t,\lambda)+ P_{\mathfrak{ m}}(x,t,\lambda).$
    The compatibility of the operators $L$ and $M$ \eqref{Lax1} gives
    $$ i Q_t = i P_{x} -[Q,P] - \lambda [J, P_{\mathfrak{ m}}]$$ where we have used $[J, P_{\mathfrak{ k}}]=0.$
      Applying $\mathfrak{ k} -\mathfrak{ m}$ decomposition, this splits further to 
 \begin{align*}
   &  i Q_t = i P_{\mathfrak{ m},x} -[Q,P_{\mathfrak{ k}}] - \lambda [J, P_{\mathfrak{ m}}],\\
  &  i P_{\mathfrak{ k},x}=[Q,P_{\mathfrak{ m}}].
\end{align*}  From the last relation we determine
    $$ P_{\mathfrak{ k}}=-i \partial ^{-1} [Q,P_{\mathfrak{ m}}]. $$
    Then $Q_t=(\partial +\mathrm{ad}_Q  \partial^{-1} \mathrm{ad}_Q +i\lambda \mathrm{ad}_J  )  P_{\mathfrak{ m}}.$
    Assuming $P=\sum_{j=0}^N  P^{(j)}\lambda ^j, $ we have for $\lambda^0$
    \begin{equation}\label{nlee}
         i Q_t = i P_{\mathfrak{ m},x}^{(0)} -[Q,P_{\mathfrak{ k}}^{(0)}] = i P_{\mathfrak{ m},x}^{(0)} +i[Q,\partial ^{-1} [Q,P_{\mathfrak{ m}}^{(0)}]]
    \end{equation}
      and for the higher powers of $\lambda$
 \begin{align*}
   (\partial +\mathrm{ad}_Q  \partial^{-1} \mathrm{ad}_Q) P_{\mathfrak{ m}} ^{(j)}= -i \mathrm{ad}_J P_{\mathfrak{ m}} ^{(j-1)}, \quad j= 1,...,N+1,
\end{align*}
     
             When $N=2$ for example, $[J,P_{\mathfrak{ m}} ^{(2)} ]=0$ and we take $P_{\mathfrak{ m}} ^{(2)}=0.$
        One can take $P_{\mathfrak{ k}} ^{(2)}=J.$
    Then $[J,P_{\mathfrak{ m}} ^{(1)} ]= [P_{\mathfrak{ k}} ^{(2)},Q]=[J,Q]$ then $P_{\mathfrak{ m}} ^{(1)}=Q$ and 
    $P_{\mathfrak{ k}} ^{(1)}=0.$  Furthermore,  $i P_{\mathfrak{ m},x} ^{(1)}=[J, P_{\mathfrak{ m}} ^{(0)}]$ will provide $P_{\mathfrak{ m}} ^{(0)}$ in terms of $Q,$ \begin{equation}
    P_{\mathfrak{ m}} ^{(0)}=i \text{ad}_J^{-1}Q_x.
\end{equation}  The nonlinear evolution equation \eqref{nlee} can be written down also as 
\begin{equation}
    iQ_t+\text{ad}_J^{-1}Q_{xx}+[Q, \partial^{-1}[Q, \text{ad}_J^{-1}Q_x]]=0,
\end{equation}
which represents the generalisation of the NLS. The invertibility of $\text{ad}_J$ on the symmetric space $\mathfrak{m}$ could be illustrated explicitly as follows.  The variable $Q \in \mathfrak{m}$ admits a decomposition over the Cartan-Weyl basis $\{E_{\pm \alpha}, \alpha \in \theta^+ \}$ ($\theta^+$ is a subspace of the root space of $\mathfrak{g}$) spanning $\mathfrak{ m}:$ 
    $$Q= \sum_{\alpha \in \theta^+} (q^{\alpha}E_{\alpha}+p^{\alpha} E_{-\alpha}).$$
        Since  $Q_x=P_{\mathfrak{ m},x} ^{(1)}=[J,P_{\mathfrak{ m}} ^{(0)}] $ and
    $\mathrm{ad}_J( E_{\pm \alpha})=\pm a E_{\pm \alpha}$ when $\alpha \in \theta^+$ then
    $$P_{\mathfrak{ m}} ^{(0)} = \frac{1}{a}  \sum_{\alpha \in \theta^+} (q_x^{\alpha}E_{\alpha}-p_x^{\alpha} E_{-\alpha})$$ could be obtained explicitly.
    
Let us present an example with the symmetric space {\bf A.III} (see \cite{KuFo,VSG-basic}), which bears also the notation 

$ SU(m+n)/S(U(m)\times U(n)).$  The complexification of the algebra $su(n, \mathbb{R})$ is isomorphic to $sl(n, \mathbb{C}) \sim A_{n-1}.$ 
 Let us specify $ m=n=2,$   $ J=\frac{1}{2}\mathrm{diag}(1,1,-1,-1)$ and
$$Q= \left(\begin{array}{cccc} 
0 & 0 & q_1 & q_2 \\
0 & 0 & q_4 & q_3 \\
-\bar{q}_1 & -\bar{q}_4 & 0 & 0 \\
-\bar{q}_2 & -\bar{q}_3 & 0 & 0
\end{array}\right) .$$
        
This parameterization assumes $p^{\alpha}=-\bar{q}^{\alpha}$. The equations written for each component are
\begin{align*}
   & i q_{1,t}= q_{1,xx}+2q_1(|q_1|^2 + |q_2|^2 +|q_4|^2 ) +2 q_2 q_4 \bar{q}_3                  \\
      & i q_{2,t}= q_{2,xx}+2q_1(|q_2|^2 + |q_2|^2 +|q_3|^2 ) +2 q_1 q_3 \bar{q}_4                  \\
   & 1\leftrightarrow 3, \quad  2\leftrightarrow 4.
\end{align*}
    
    In general, for the  symmetric space {\bf A.III}, the matrix  notations are very useful:
         $ J=\frac{1}{2}\mathrm{diag}(\mathbf{1}_n,-\mathbf{1}_m)$ and
$$Q= \left(\begin{array}{cc} 
0 & \q \\
\p & 0 
\end{array}\right)$$ where $\mathbf{1}_n$ is the  $n\times n$ unit matrix, $\q$ and $\p^T$ are matrices $n\times m$ ($\q$ is a vector if $n=1.$)
        
    The equations, written in matrix form are 
\begin{equation}\label{eq:mNLEE2}\begin{split}
i \frac{\partial \q}{ \partial t} &=\frac{\partial^2 \q }{ \partial x^2 } +2 \q \p \q , \\
-i \frac{\partial \p}{ \partial t} &= \frac{\partial^2 \p }{ \partial x^2 } + 2 \p \q \p .
\end{split}\end{equation}
 Further reduction $\p=\pm\q^{\dagger}$ leads to 
    $$i \frac{\partial \q}{ \partial t} =\frac{\partial^2 \q }{ \partial x^2 } \pm 2 \q \q^{\dagger} \q.$$
    If $\q$ is a vector ($n=1$) this is the vector NLS (Manakov's equation), $\q \q^{\dagger}=||\q||^2$ is the norm of the vector.

    Another possible reduction which recently gained popularity is $\p(x,t)=\pm\q^{T}(-x,-t),$ leading to the non-local NLEE
 $$i \frac{\partial \q}{ \partial t} =\frac{\partial^2 \q }{ \partial x^2 } \pm 2 \q \q^{T}(-x,-t) \q.$$
    A nicely written review article \cite{VSG-basic} introduces the basic features of the integrable equations on symmetric spaces. It describes the fundamental properties and underlying structures of the nonlinear evolution equations (NLEE), based on the expansions of $Q, Q_t$ over the squared eigenfunctions of $L$ \eqref{Lax1} (which are eigenfunctions of the recursion operator), the hierarchy of NLEE, the hierarchy of integrals of motion and the hierarchy of Hamiltonian structures. The completeness of the squared eigenfunctions is of fundamental significance, since it makes the Inverse Scattering Transform a generalised Fourier Transform. For an $sl(2)$-valued Lax operator the completeness has been formulated by D.J. Kaup \cite{Ka76}, then proven rigorously by V Gerdjikov and E. Hristov \cite{GH1,GH2}. There the
authors introduced also the symplectic basis, which maps the potential $Q(x, t)$ onto the action-angle variables of NLS. A general result, for $Q, J \in \mathfrak{g}$,  $J\in \mathfrak{h}$ for a semisimple $\mathfrak{g}$ has been formulated and proven by V. Gerdjikov, in \cite{VSG1}. 

For systems on symmetric spaces, as well as for systems with $\mathbb{Z}_h$ - Mikhailov reductions \cite{Mi} the expansions over the squared eigenfunctions have been derived in \cite{VSG-basic,VSG2,VSG3,GSM,GY}. As a result, the spectral theory of the Lax operators has been described and the arising hierarchies of Hamiltonian structures have been studied.

\section{Quadratic pencil of Lax operators }

Lax operators, for which $L$ is quadratic in the spectral parameter $\lambda$ also lead to
multicomponent integrable systems. These systems generalise the family of the DNLS equations - DNLS I, or Kaup-Newell equation, \cite{KN, GIK80, For} , DNLS~II \cite{ChLi} and DNLS~III, or Gerdjikov-Ivanov equation, \cite{GI0, GI}, the Fokas-Lenells equation, \cite{LF} and others. 
Now we illustrate these possibilities for $M-$operator representing ''positive'', ''negative'' and 
''rational'' flows \cite{GGI,V13,V16,GI21,VSG2,RI22}. 

\subsection{"Positive" flows: DNLS equations on symmetric spaces}

Let us illustrate this possibility again with the {\bf A.III} Hermitian symmetric space, $SU(m+n)/(S(U(m)\otimes U(n))$.  
The Lax operators are given by \cite{GGI}:
\begin{equation}\label{eq:L2}\begin{split}
 L \psi &\equiv i \frac{\partial \psi }{ \partial x } + (U_2(x,t) + \lambda Q(x,t) -\lambda^2 J) \psi(x,t,\lambda) =0,\qquad
  Q(x,t) = \left(\begin{array}{cc} 0 & \q \\ \p & 0   \end{array}\right), \\
 M \psi &\equiv \!i \frac{\partial \psi }{ \partial t }\! +\! (V_4(x,t) \!+ \!\lambda V_3(x,t) + \lambda^2 V_2(x,t)
 +  \lambda^3 Q(x,t)-\lambda^4 J) \psi=0.
\end{split}\end{equation}
where $Q(x,t)$,  $   V_3(x,t) \in \mathfrak{m}$ and $U_2(x,t)$,  $ V_2(x,t)$ and  $ V_4(x,t) \in \mathfrak{k}$, 
$\q$ is a $n\times m$ matrix, $\p$ is a $m\times n$ matrix.

Such Lax pairs give rise to multicomponent derivative NLS type equations. Indeed, the Lax pair (\ref{eq:L2}), leads to the system of NLEE, generalising the DNLS III (V. Gerdjikov - M. Ivanov, \cite{GI,GI0}) equation:
\begin{equation}\label{eq:NLEE2*}\begin{split}
i \frac{\partial \q}{ \partial t} + \frac{1}{2} \frac{\partial^2 \q }{ \partial x^2 } - \frac{i}{2} \q \frac{\partial \p}{ \partial x } \q + \frac{1}{4} \q \p \q \p \q &=0, \\
-i \frac{\partial \p}{ \partial t} + \frac{1}{2} \frac{\partial^2 \p }{ \partial x^2 } + \frac{i}{2} \p \frac{\partial \q}{ \partial x } \p + \frac{1}{4} \p \q \p \q \p &=0.
\end{split}\end{equation}

Reductions, relating $\p$ and $\q$ are possible, such as $\p=\pm \q^{\dagger}$ and   $\p(x,t)=\pm \q^{T}(-x,-t).$

Along with  \eqref{eq:L2} one can consider the Lax pair \cite{GGI}:
\begin{equation}\label{eq:L2t}\begin{split}
\tilde{L} \tilde{ \psi} &\equiv i \frac{\partial \tilde{ \psi} }{ \partial x } +  (\lambda \tilde{ Q}(x,t) -\lambda^2 J)\tilde{ \psi}(x,t,\lambda) =0, \quad \tilde{ Q}(x,t) = \left(\begin{array}{cc} 0 & \tilde{ \q} \\ \tilde{ \p} & 0   \end{array}\right), \\
 \tilde{ M} \tilde{ \psi} & \equiv  i \frac{\partial \tilde{\psi} }{ \partial t } + ( \lambda \tilde{V}_3(x,t) + \lambda^2 \tilde{V}_2(x,t)
+  \lambda^3 \tilde{Q}(x,t) -\lambda^4 J)\tilde{ \psi}(x,t,\lambda)=0.
\end{split}\end{equation}

This Lax pair is gauge equivalent to the previous one (\ref{eq:L2}). The corresponding system is generalising the Kaup-Newell equation \cite{KN}:
\begin{equation}\label{eq:KNm*}\begin{split}
i \frac{\partial \tilde{\q}}{ \partial t } &+ \frac{\partial^2 \tilde{\q} }{ \partial x^2 } + i \frac{\partial \tilde{\q}\tilde{\p}\tilde{\q}}{ \partial x }=0, \\
-i \frac{\partial \tilde{\p}}{ \partial t } &+ \frac{\partial^2 \tilde{\p} }{ \partial x^2 } - i \frac{\partial \tilde{\p}\tilde{\q}\tilde{\p}}{ \partial x }=0.
\end{split}\end{equation}
This system also complies with the grading introduced by the symmetric space:
$\tilde{ Q}(x,t)$,  $   \tilde{ V}_3(x,t) \in \mathfrak{m}$ and $\tilde{ V}_2(x,t) \in \mathfrak{k}$.

Reductions, relating $\tilde{\p}$ and $\tilde{\q}$ are possible, such as $\tilde{\p}=\pm \tilde{\q}^{\dagger}$ and $\tilde{\p}(x,t)=\pm \tilde{\q}^{T}(-x,-t)$. 
Other examples of the generalised KN system, related to the {\bf A.III} and {\bf BD.I} symmetric spaces are presented in \cite{V13,V16}.

\subsection{Negative flows: Fokas-Lenells  (FL) equation on symmetric spaces}

The Fokas-Lenells equations (see \cite{GI21} for details) are associated to the so-called {\it negative flows} and the following Lax pair
\begin{equation}\label{eq:Lax}\begin{split}
& i \Psi _x + (\lambda Q_x - \lambda^2 J) \Psi =0,\\
& i \Psi_t +\left(\lambda Q_x  + V_0 + \lambda^{-1} V_{-1} -(\lambda^2-\frac{2}{a}+ \frac{1}{a^2\lambda^{2}}) J\right) \Psi,
\end{split}\end{equation} 
\begin{equation}\label{eq:Q}
\text{where} \quad Q(x,t)=\sum_{\alpha \in \theta^+} (q^{\alpha} E_{\alpha}+p^{\alpha} E_{-\alpha}) \in \mathfrak{m}.
\end{equation}
From the compatibility condition of the Lax operators one  can determine
\begin{equation}\label{eq:V0Vm1}
 V_{-1}=\frac{i}{a} \sum_{\alpha \in \theta^+} (q^{\alpha} E_{\alpha}-p^{\alpha} E_{-\alpha})  \in \mathfrak{m}  , \qquad
 V_0  =\frac{1}{a} \sum_{\alpha, \beta \in \theta^+} q^{\alpha}p^{\beta}[ E_{\alpha}, E_{-\beta}]  \in \mathfrak{k}.
\end{equation}

As an example we take the {\bf A.III} Hermitian symmetric space, $SU(m+n)/(S(U(m)\otimes U(n))$, with  $J$ and $ Q $ taken as matrices in the form
\begin{equation}\label{eq:A3mJQ}\begin{split}
 J = \frac{1}{m+n} \left(\begin{array}{cc} n \mathbf{1}_m & 0 \\ 0 & -m\mathbf{1}_n \end{array}\right) \in \mathfrak{k}, \quad   Q(x,t) = \left(\begin{array}{cc} 0 &  \q \\  \p  &  0  \end{array}\right) \in \mathfrak{m}
\end{split}\end{equation} 
and 
\begin{equation}\label{eq:SpV}\begin{split}
V_{-1}(x,t) = i \left(\begin{array}{cc} 0 &  \q \\  -\p  &  0  \end{array}\right), \quad
 V_0(x,t) =\left(\begin{array}{cc} \q \p & 0 \\ 0 & -\p \q   \end{array}\right).
\end{split}\end{equation}
 The equations in block-matrix form are
 \begin{equation}\label{eq:p-q_A3m}\begin{aligned}
& i\left( \q_{xt}- \q_{xx} +\q \right)   + 2 \q_x + (\q_{x} \p \q + \q \p \q_{x} ) =0  , \\
& i\left( \p_{xt}- \p_{xx} + \p \right) -2\p_x  - (\p_{x} \q \p + \p \q \p_{x})  =0.
\end{aligned}\end{equation}
Introducing new matrices $\u,\v$ such that
\[\q=e^{-ix}\u, \qquad \p=e^{ix}\v \] we represent the equations \eqref{eq:p-q_A3m}
in the form
\begin{equation}\label{eq:u-v_A3m}\begin{aligned}
 i\u_t -\u_{xt}+ \u_{xx}  + (\u+i\u_{x})\v\u+ \u\v (\u+i\u_x)=&0  , \\
 -i \v_t - \v_{xt}+ \v_{xx} +(\v-i\v_{x}) \u \v + \v \u (\v-i\v_{x})  =&0.
\end{aligned}\end{equation}
The first reduction involves Hermitian conjugation $\v=\pm \u^{\dagger},$  the equations \eqref{eq:u-v_A3m}  reduce to
 \begin{equation}\label{eq:R1_A3}
 i\u_t -\u_{xt}+ \u_{xx}  \pm (2\u\u^{\dagger}\u+i\u_{x}\u^{\dagger}\u + i\u\u^{\dagger} \u_x)=0.
\end{equation}
The second reduction $\v(x,t)=\pm \u^{T}(-x,-t),$ leads to the following nonlocal equation
\begin{equation}\label{eq:R2_A3}
 i\u_t -\u_{xt}+ \u_{xx}  \pm (2\u \tilde{\u} \u +i\u_{x}\tilde{\u}\u+ i\u \tilde{\u}\u_x)=0,
  \end{equation}
where $\tilde{\u}={\u}^{T}(-x,-t).$

\subsection{Quadratic pencil - Rational flows }

The spectral problem is quadratic with respect to the spectral parameter $\lambda,$
\begin{equation}\label{eq:LA3}\begin{split}
 i \frac{\partial \psi}{ \partial x }& =  L\psi =(\lambda^2 J + \lambda Q + P) \psi(x,t,\lambda), \\
i \frac{\partial \psi}{ \partial t }& =  M\psi = \frac{1}{\lambda^2 - \zeta^2} \left(\lambda^2 J+\lambda U + W \right)  \psi(x,t,\lambda) ,
\end{split}\end{equation} where $L,M \in \mathfrak{g},$ $P,W \in \mathfrak{k}$ and $Q,U \in \mathfrak{m}.$ The details could be found in \cite{RI22}.

The matrix realisation of the symmetric spaces is with block matrices, such that the splitting \eqref{spl} is related to the matrix block structure for the corresponding symmetric space. For the {\bf A.III} symmetric space the block structure is

\begin{equation}\label{eq:LA4}\begin{split}
 i \frac{\partial \psi}{ \partial x }& =  L\psi =\left(\begin{array}{cc} \frac{1}{2}\lambda^2 \mathbf{1} +\p_1(x,t) & \lambda \q \\ \lambda \r & -\frac{1}{2}\lambda^2\mathbf{1} +\p_2(x,t)  \end{array}\right) \psi(x,t,\lambda), \\
i \frac{\partial \psi}{ \partial t }& =  M\psi = \frac{1}{\lambda^2 \!- \!\zeta^2}\! \left(\begin{array}{cc} \frac{1}{2}\lambda^2 \mathbf{1} +\w_1(x,t) & \lambda \u \\ \lambda \v& -\frac{1}{2}\lambda^2 \mathbf{1} +\w_2(x,t)  \end{array}\right)  \psi ,
\end{split}\end{equation}
where $\q,\r,\u,\v,\p_1,\p_2,\w_1,\w_2$ are matrices of corresponding dimensions, $\zeta$ is a constant.

The compatibility condition and the change of the variables $x$ and $t$ gives
\begin{equation}\label{eq:A33}\begin{aligned}
& i\w_{1,t}-\zeta^2 i (\q\r)_x+[\q\r, \w_1]=0, \\
-& i\w_{2,t}-\zeta^2 i (\r\q)_x+[\r\q, \w_2]=0, \\
& i\q_t +\!\q_{xt}\!+\zeta^2 i\q_x +\q\r(\q-i\q_x)+(\q-i\q_x)\r\q +\w_1\q-\q\w_2=0,\\
-& i\r_t +\!\r_{xt}\! -\zeta^2 i\r_x+(\r+i\r_x)\q\r+\r\q(\r+i\r_x)+\r\w_1 -\w_2\r=0.
\end{aligned}\end{equation}

The reduction $\r=\q^{\dagger}$ leads to $\w_1=\w_1^{\dagger},$ $\w_2=\w_2^{\dagger},$ and the coupled system of equations
 \begin{equation}\nonumber
 \begin{aligned}
& i\w_{1,t}-\zeta^2 i (\q\q^{\dagger})_x+[\q\q^{\dagger}, \w_1]=0, \\
-& i\w_{2,t}-\zeta^2 i (\q^{\dagger}\q)_x+[\q^{\dagger}\q, \w_2]=0, \\
& \!i\q_t +\!\q_{xt}+\!\zeta^2\! i\q_x \!+\!\q\q^{\dagger}(\q-\!i\q_x)+(\q-\!i\q_x)\q^{\dagger}\q\! +\w_1\q-\!\q\w_2\!=\!0.\\
\end{aligned}\end{equation}

 The reduction $\r=-\q^{\dagger}$ is also possible.
If $q=\q$ is just a scalar we have the one-component integrable equation 
\begin{equation}\label{eq:Q3r1}
 iq_t +\zeta^2 iq_x+q_{xt}\pm 2|q|^2(q-iq_x)\mp 2  \zeta^2 q \partial_t^{-1}(|q|^2)_x=0 .
\end{equation}

The integration operator $ \partial_t^{-1}$ could be understood as $\int_{-\infty}^{t} dt',$ which leads to a hysteresis term in the equation.  We assume for simplicity that all functions are from the Schwartz class $\mathcal{S}(\mathbb{R})$ in $x$ for all values of $t.$ A non-local reduction exists, $r(x,t)=\pm q(-x,-t),$ (scalar functions). The equation is (only the $-x$ and $-t$ arguments are explicit)
\begin{equation}\label{ewh}
i q_t+ \zeta^2 iq_x +q_{xt}\pm 2q q(\!-\!x,\!-\!t)(q-iq_x) \mp 2  \zeta^2 q \int_{-\infty}^{t}\big(q q(\!-\!x,\!-\!t)\big)_x d t'=0 .
\end{equation}
One possible interpretation of \eqref{ewh} is a nonlocal (due to the "$-x$" dependence) non-evolutionary equation with hysteresis.

The inverse scattering for the spectral problem of the quadratic bundle of $sl(2, \mathbb{C})-$valued $L$-operators has been described in \cite{GIK80}. The normalization of the associated RHP problem is canonical which facilitates the computations leading to the soliton solutions.

\section{Modelling with higher order nonlinear Schr\"{o}dinger equations }
 From modelling point of view, the generalised NLS or Higher order NLS (HNLS) equations with applications in nonlinear optics \cite{Ko,KoHa} as well in water waves \cite{GH} and plasma \cite{GT96} are usually written in the form
 
 \begin{equation}\label{NLStype}
  iq_{T} +i c  q_X+\frac{1}{2}q_{XX} + |q|^2q+i\beta_1q_{XXX}+i\beta_2|q|^2q_X+i\beta_3 q (|q|^2)_X=0.
\end{equation}
where $c,$ $\beta_i,$ $i=1,2,3$ are, in general arbitrary real constants, depending on the physical parameters. 

The integrable cases correspond to the following ratios $(\beta_1: \beta_2:\beta_3):$  The DNLS I and II with $(0:1:1)$ and $(0:1:0)$; Hirota \cite{Hi} - with $(1:6:0)$ and Sasa-Satsuma \cite{SS91} - with $(1:6:3).$
Nijhof and Roelofs \cite{NR92} using the prolongation method have proven that no other integrable cases in the form \eqref{NLStype} exist.

For practical applications, however, there is usually a small parameter, $\varepsilon,$ such that the quantities scale like $q\sim \varepsilon , $ i.e. $q\rightarrow \varepsilon q,$ $T=\varepsilon t,$ slow time, then $t=T/\varepsilon$ and $X=\varepsilon x$ - slow space variable, where $(x,t)$ are the original unscaled variables. Therefore, the physical models leading to HNLS involve perturbative expansions like
\begin{equation}\label{NLStype1}
  i q_{t} + i c q_x+\frac{\varepsilon}{2}q_{xx} + \varepsilon |q|^2q+i\varepsilon^2\!\left(\beta_1q_{xxx}\!+\!\beta_2|q|^2q_x +\beta_3 q (|q|^2)_x\right) =\mathcal{O}(\varepsilon^3).
\end{equation}
 Thus we can extend the set of integrable models by considering those, which admit the scaling as in \eqref{NLStype1}, and we can use the triples $(\beta_1:\beta_2:\beta_3)$ for some systematic classification of the models.

 For example, the Fokas - Lenells equation \cite{LF}, 
 \begin{equation}\label{LF}
  i q_t +i c q_x-\varepsilon \nu q_{xt} + \varepsilon  \gamma q_{xx} + \varepsilon |q|^2 q + i\varepsilon^2 \nu |q|^2 q_x=0,
\end{equation}
where $c, \nu, \gamma$ are constants, can be transformed as follows. In the leading order, $q_t=-c q_x$ thus we have
\begin{equation}\label{LF0}
\begin{split}
  &i q_t +i c q_x +\varepsilon (\gamma+\nu c) q_{xx} +  \varepsilon |q|^2 q =\mathcal{O}(\varepsilon^2),\\
  &q_t=-c q_x+i\varepsilon (\gamma+\nu c) q_{xx} + i \varepsilon |q|^2 q +\mathcal{O}(\varepsilon^2)
  \end{split}
\end{equation}

Next we substitute $q_t$ from \eqref{LF0} in the $q_{xt}$ term of \eqref{LF} to obtain
\begin{equation}\label{LF1}
 i q_t +i c q_x+\varepsilon (c\nu + \gamma) q_{xx} +\! \varepsilon |q|^2 q -i\varepsilon^2 \nu( c \nu\!+\!\gamma)q_{xxx}- i\varepsilon^2 \nu q( |q|^2 )_x= \mathcal{O}(\varepsilon ^3).
\end{equation}

In order to match \eqref{NLStype} we need to choose $c\nu+\gamma=1/2,$ then

\begin{equation}\nonumber
 i q_t +i c q_x+ \frac{\varepsilon }{2} q_{xx} + \varepsilon |q|^2 q -i\varepsilon^2 \frac{\nu}{2}\big( q_{xxx} +2 q( |q|^2 )_x\big)=\mathcal{O}(\varepsilon ^3),
\end{equation}
hence the model is characterised by the ratio $(1:0:2).$
The non-evolutional equation \eqref{eq:Q3r1}
\begin{equation}\nonumber
 iq_t +\zeta^2 iq_x+q_{xt}\pm 2|q|^2(q-iq_x)\mp 2  \zeta^2 q \partial_t^{-1}(|q|^2)_x=0 
\end{equation}
with the upper sign is characterised by $(1,-1 ,3)$ and $\zeta^2=-2;$ with the lower sign - by $(1,-2 ,6)$ and $\zeta^2=2.$

For matching physical models \eqref{NLStype1}  with arbitrary coefficients $\beta_1,\beta_2,\beta_2$  to  an integrable equation, one needs to apply the method of Kodama transformations. For HNLS they are provided in \cite{Ko85}.

\section{Conclusions}
 The symmetric spaces approach provides a convenient setting for constructing and classifying multi-component integrable systems. The negative and rational flows lead to a number of new integrable nonlinear systems in non-evolutionary form.

The spectral theory, inverse scattering, recursion operators, completeness of squared eigenfunctions, hierarchies etc. remain to be studied in detail for the quadratic pencil of Lax operators. In particular, the completeness
of the squared eigenfunctions of a given Lax operator is a powerful tool for the description of whole hierarchy of NLEE and their properties, so the negative flows outlined above as well as other members of the hierarchy should come out from such description.

 From the modelling point of view, the asymptotic expansions are important. Nonevolutionary integrable equations could be asymptotically equivalent to evolutionary non-integrable (but physically important) equations.

\begin{acknowledgments}
The author is thankful to Prof. V. Gerdjikov and Dr G. Grahovski for their help and advice. Partial funding from grant 21/FFP-A/9150 (Science Foundation Ireland) and grant K$\Pi$-06H42/2 (Bulgarian National Research Foundation) is gratefully acknowledged.

\end{acknowledgments}

\nocite{*}
\bibliography{aipsamp}

\end{document}